\documentclass[a4paper,11pt]{scrartcl}
\usepackage[margin=1in]{geometry} 

\usepackage{avant}

\usepackage{setspace}
\renewcommand{\contentsname}{Overview}
\defcaptionname{\languagename}{\contentsname}{Overview}

\usepackage[utf8]{inputenc}

\usepackage{url}

\usepackage{breakurl}
\usepackage[breaklinks]{hyperref}

\usepackage[style=numeric-comp, sorting=none, backend=biber, useprefix, doi=true, isbn=true, giveninits=true, uniquename=false, maxbibnames=99, maxcitenames=2]{biblatex}
\title{Priorities for more effective tech regulation}
\author{Konrad Kollnig}

\date{
	Department of Computer Science, University of Oxford
}

\begin{document}
	\maketitle
	
        \noindent Ample research has demonstrated that compliance with data protection principles remains limited on the web and mobile~\cite{kollnig2021_consent,kollnig2022_iphone_android,kollnig_before_2021,reyes_wont_2018,maps_2019,okoyomon_ridiculousness_2019,nguyen_share_first_consent_2021,nouwens_dark_2020,masse_two_2020,Veale_Nouwens_Santos_2022}.
        For example, almost none of the apps on the Google Play Store fulfil the minimum requirements regarding consent under EU and UK law, while most of them share tracking data with companies like Google/Alphabet and Facebook/Meta and would likely need to seek consent from their users~\cite{kollnig2021_consent}.
        Indeed, recent privacy efforts and enforcement by Apple have had~--~in some regards~--~a more pronounced effect on apps' data practices than the EU's ambitious General Data Protection Regulation (GDPR)~\cite{kollnig_att_2022,kollnig_cost_2022}.
        Given the current mismatch between the law on the books and data practices in reality, iterative changes to current legal practice will not be enough to meaningfully tame egregious data practices. Hence, this \textit{technical report} proposes a range of priorities for academia, regulators and the interested public in order to move beyond the status quo.
        This work is an excerpt from my PhD dissertation at the University of Oxford under the supervision of Sir Nigel Shadbolt.


    \linespread{1}
	\tableofcontents
    \linespread{1.5}
    \newpage

\section*{Priority 1: Make consent \emph{meaningful}~--~or abandon it}\addcontentsline{toc}{section}{Priority 1: Make consent \emph{meaningful}~--~or abandon it}

Although often claimed otherwise, the GDPR \emph{does not} require a broad implementation of `cookie banners'. EU and UK data protection principles have hardly changed since the GDPR came into force in May 2018 and were already part of the Data Protection Directive 1995. The requirements regarding `cookie banners' additionally result from Art 5(3) of the ePrivacy Directive as amended in 2009, not from the GDPR.

The recent flood of cookie banners can rather be explained by the fact that the potential sanctions for data protection violations have drastically increased with the GDPR, causing discontent within the online data industry. The conditions for consent have also been tightened and existing standards have been clarified in line with case law. User consent must now be `freely given, specific, informed and unambiguous' (Recital 32 GDPR). However, this is rarely the case in practice~\cite{nouwens_dark_2020,matte_cookie_2019,kollnig2021_consent,nguyen_share_first_consent_2021}. A significant proportion of current `cookie banners' are thus in violation of the GDPR.

The designation of those \emph{consent} banners as `cookie banners' can further be interpreted as misinformation. For example, Facebook implements a pop-up on its website titled `Allow the use of cookies from Facebook in this browser?' It is only in the accompanying Cookie Policy that Facebook clarifies that `cookies' do not only refer to cookies but also that other `technologies, including data that we store on your web browser or device, identifiers associated with your device and other software, are used for similar purposes.' The online advertising industry today uses a variety of technologies to track user activity across various apps and websites~--~such as fingerprinting (i.e.~using browser characteristics such as time zone, language and operating system) and email hashing (i.e.~sending email addresses from non-Facebook websites to Facebook even if the user does not use Facebook). This collection of data about websites and apps~--~tracking~--~is widespread, as found by numerous pieces of research~\cite{binns_third_2018,kollnig2022_iphone_android,razaghpanah_apps_2018}. Meanwhile, the term `cookie' sounds innocuous and is widely used by the industry.

Overall, a considerable part of the `cookie banners' on the internet aims to misinform and frustrate internet users vis-à-vis the GDPR rather than to implement the law's requirements~\cite{Veale_Nouwens_Santos_2022}.

There remains significant work to do for authorities and other organisations to tackle incompliant implementations of consent and make it meaningful. Indeed, ample research suggests that this is not possible at all~\cite{solove_privacy_2012,barocas2009notice,bietti_consent_2020}, in part because individuals will never be sufficiently `informed'~--~as is required by GDPR~--~about the opaque data practices of large technology companies~\cite{norwegian_consumer_council_out_2020}.

\section*{Priority 2: Better, bolder communication}\addcontentsline{toc}{section}{Priority 2: Better, bolder communication}

Due to the continued uncertainty and misinformation regarding the GDPR, the current way of working of data protection and other public authorities has created a vacuum of knowledge and authority that has been successfully occupied by third parties with strong self-interests. This way of working in the EU is often characterised as bureaucratic and apolitical, resulting from a lack of a transnational public sphere in Europe. However, without a European public sphere and debate, political legitimacy in the Habermasian sense is difficult, if not impossible.

The end result is problematic for data protection because it fuels a negative and dismissive mood among citizens~--~including those individuals who are responsible for the practical implementation of the GDPR~--~towards the competence of public authorities in digital matters. Data protection and other public authorities should counter this perception boldly and decisively. This applies both to new digital initiatives and existing laws such as the GDPR.

\section*{Priority 3: Clear technical standards, visualisations and reference code}\addcontentsline{toc}{section}{Priority 3: Clear technical standards, visualisations and reference code}

As part of better communication, \emph{clear, reliable and actionable technical standards} should be considered. Unfortunately, developers do often not know how to comply~\cite{anirudhchi2021,mhaidli_we_2019,sirur_are_2018}, so there is a need to clarify what forms of data processing are permitted and how this should be implemented in software.

Currently, the expectation from the authorities is that software developers will resolve important issues related to the implementation of the GDPR themselves~--~by studying the relevant legislation and rulings. This assumption is unrealistic, at least for smaller software companies~\cite{sirur_are_2018}. In addition, the European Data Protection Board and the ICO regularly publish explanatory notes on important aspects of the GDPR. This usually involves the publication of long texts of legalese. The target audience of these publications is thus primarily legal, especially courts, but not the individuals tasked with the practical implementation of the law.

It is certainly important to explain the legal dimensions of the GDPR and to pursue this through legal methodology, particularly by publishing explanatory legal texts. At the same time, it seems that authorities too often hide their lack of authority and technical expertise behind overly formal communication and shy away from clear specifications. As a result, a significant part of the interpretation of the GDPR currently falls to the courts. Unfortunately, this approach undermines a swift and effective implementation of the GDPR and is unsuitable to keep pace with rapid technological change. Code can be changed and rolled out to users worldwide in a matter of minutes. For effective IT regulation, the (ambitious) goal must be to act similarly agile.

From a technical perspective, it is almost naïve to assume that legal text could be translated more or less directly into code. Instead, in IT, \emph{requirements specification} provides a decades-old approach to describing and building IT systems. A common standard was first published by the IEEE (Institute of Electrical and Electronic Engineers) in 1984; the latest version is ISO/IEC/IEEE 29148:2018 from 2018. Requirements can be both technical and non-technical as well as specific and less specific. There is no reason why similar requirements cannot be formulated for core elements of the GDPR and other IT law. This could be done in particular for the implementation of consent and should be accompanied by visualisations and reference code where possible. In the context of the amendment of the ePrivacy Directive in 2009, the EU even provided visualisations and reference code in the past, but did not maintain them over the years and discontinued them after the introduction of the GDPR.

\section*{Priority 4: Sufficient resources for authorities}\addcontentsline{toc}{section}{Priority 4: Sufficient resources for authorities}

There are many reasons for the lack of implementation of the GDPR in practice. One important reason is the continued lack of resources of data protection authorities~\cite{edpb_funding,lynskey_grappling_2019}. This refers to both financial resources and (technical) expertise. For example, there has been virtually no action by the responsible authorities against the documented data protection problems in mobile apps. A second reason is the one-stop-shop approach of the GDPR in the EU. This approach currently leads to a race to the bottom between member states in terms of negligent implementation of the GDPR. In particular, Ireland, where most of the major tech companies in Europe are based (including Microsoft, Alphabet/Google and Meta/Facebook), has been criticised in this regard~\cite{mcintyre_regulating_2021,iccl_ads}. A third reason is the still-evolving case law in the courts.

The problem of the lack of practical enforcement of the GDPR has been recognised by lawmakers and is being addressed in new EU digital legislation. Ireland is no longer a single point of failure of legal enforcement in DMA and DSA against big tech, as it was de facto under the GDPR, but rather the EU Commission. In addition, technology companies will be required to subsidise enforcement financially.

A key challenge will remain recruiting the necessary technical talent for public institutions, most of whom currently work for the same technology companies and are needed to keep pace with private industry in terms of expertise and technical understanding. In the past, European legislators have not always maintained an air of technical competence. One example is the planned EU AI Act, which is supposed to regulate AI applications. However, the definition of AI applications in the Commission's first proposal was so broad that it covered almost any computer application. The planned AI rules are derived from EU product safety legislation. This creates the risk of missing the core of AI, which rather lies in the inputs and outputs of the model rather than the product/technology itself. Doubts about the EU legislator's deep technical understanding also arise when reading the GDPR. The law, like its 1995 predecessor, distinguishes between controllers and processors in the processing of personal data. Controllers are those that alone or jointly with others determine the purposes and means of the processing of personal data (Art 4 GDPR). Processors, on the other hand, usually only act at the instruction of the controller. However, today's IT systems are the product of the combined work of many different actors, both large and small. This often makes it almost impossible to distinguish between controllers and processors. This distinction is, however, important because controllers face many more obligations than processors. Moreover, whether and to what degree software development~--~rather than direct data processing~--~entails obligations under the GDPR is not clear~\cite{bygrave_data_2017,jasmontaite_data_2018}. As a result of these definitions, which only peripherally deal with the usual processes and distribution of tasks in software development, there are a number of concluded and pending cases regarding the definition of the role of data controller~\cite{ecj_waka,ecj_jw,ecj_fashion,belgium_dpa_tcf}. One solution to this phenomenon was proposed by the Belgian DPA~\cite{belgium_dpa_tcf} in its case against IAB Europe: the DPA decided to define almost all actors in the online advertising business as controllers, i.e.~thousands of different companies. IAB Europe has appealed and the case is currently pending before the ECJ. As an alternative approach, China's Personal Information Protection Law (PIPL) from 2021 only foresees processors of personal data, but no controllers.

Of course, the GDPR is not limited to IT but covers many other areas of our daily lives that involve personal data. Therefore, one could argue that criticism of the GDPR's lack of focus on software development misses the point of the law. However, it is also the case that without technological developments, there would have been little motivation for a revision of EU data protection law (see also Recital 6 GDPR).

\section*{Priority 5: Embrace regulatory technologies}\addcontentsline{toc}{section}{Priority 5: Embrace regulatory technologies}

There are two dominant approaches to enforcing data protection rules in digital systems. The first one is taken by data protection authorities who tend to focus their efforts on a few select cases and companies. The hope is that this will tame the most egregious data practices and that there will be spillover effects across the data practices by other organisations. The second approach is taken by gatekeepers, such as app stores, who conduct some enforcement of data protection rules at scale (e.g.~through their (automated) app review), but publish limited public information about this enforcement, including the \emph{number and nature} of decisions taken~\cite{hoboken2021}. Given the scale of the digital ecosystem and the extent of current violations of data protection rules (as observed in this and other work), both approaches are insufficient. Without the help of regulatory technologies in ensuring compliance in the digital ecosystem, it will be \emph{impossible} to scale operations across the vastness of these digital ecosystems, to fulfil the expectations of individuals in keeping them safe online, and to protect fundamental rights and freedoms.

In the app ecosystem, an important, persisting issue that emerged from my analysis across iOS and Android is the lack of transparency around apps' data practices. This conflicts with the strict transparency requirements for the processing of personal data laid out in the GDPR. Design decisions by Apple and Google currently impede research efforts, such as the application of copyright protection to \emph{every} iOS app~--~even free ones. This is why it is \emph{important to develop and maintain transparency tools}.
A starting point could be expanding my PlatformControl toolkit (\url{https://platformcontrol.org}), and give more up-to-date and detailed insights into apps' privacy and compliance properties. 
As part of this, an important field for further study is the development of a cross-platform app instrumentation tool. \emph{Automatic app compliance analysis tools} are not widely available nor used by regulators and the interested public (though it might be easy to conduct such automatic checks if the regulators defined more explicit rules regarding privacy and app design), but would help keep up with the vastness of the app ecosystem. Such analysis tools would require \emph{reliable and computable metrics for compliance}. While most of this work has been on the situation in Europe, there have been emerging many promising new pieces of technology regulation across the globe, which need further investigation.

\section*{Priority 6: Evolve `legacy' legislation and provide support for research}\addcontentsline{toc}{section}{Priority 6: Evolve `legacy' legislation and provide support for research}

Much research efforts have been devoted to analysing privacy in mobile apps. Such research remains challenging, as the creation of the necessary data is associated with high investments of time and scarce technical expertise~\cite{kollnig_ready_2023}. The fact that analysing privacy issues in apps and in other software products is so difficult has an impact not only on my research but also on the work of other researchers and data protection authorities aiming to protect fundamental rights in digital systems~\cite{kollnig_ready_2023}. For example, most data protection authorities themselves currently do not possess independent expertise to analyse compliance issues in mobile apps.

The EU Digital Services Act makes promising progress in supporting research in relation to online platforms and search engines. Its Article 40, for example, obliges `very large online platforms' and `very large online search engines' to allow researchers to analyse `systemic risks'. The concrete implications for research practice, however, remain to be seen. I have, in the context of app research, elaborated on these new legal requirements in a recent pre-print~\cite{kollnig_ready_2023}. It can, however, be expected that clarifications of the law by the highest courts will be necessary and that many years will pass before the law will lead to major changes to the status quo.

Despite all the debates about new IT laws, one must not lose sight of existing laws, such as copyright, patent, and IT security law. Even if such legislation may be less attractive for public and academic debate and thus receives less attention, there is also a great need for improvement here. This was also demonstrated by the research in my PhD thesis, which conducted the first large-scale study into iOS app privacy in about 10 years and avoided legal challenges around Apple's application of DRM to iOS apps~\cite{kollnig2022_iphone_android,kollnig_att_2022}.
The used methods are freely available at \url{https://platformcontrol.org}.
	
\paragraph{Acknowledgements}
This report is inspired by previous submissions to the UK competition authority~\cite{kollnig_cma_2022} and the Department for Digital, Culture, Media and Sport~\cite{researchers_from_the_human_centred_computing_research_group_department_of_computer_science_university_of_oxford_response_2021} from my research group. An earlier version of this report was published in the Ad Legendum journal of the University Münster (in German)~\cite{kollnig_lehren_2023}.
	

    {\renewcommand*\MakeUppercase[1]{#1}%

    \begingroup
    \sloppy
    \printbibliography
    \endgroup
	
	
	
	
\end{document}